\documentstyle[preprint,aps]{revtex}
\input{psfig}
\begin{document}
\preprint{SBRHI-99-1}

\tightenlines
\title{Light Fragment Yields from Central Au + Au Collisions at 11.5 A GeV/c}

\author{J.~Barrette$^5$, R.~Bellwied$^{9}$, S.~Bennett$^{9}$, R.~Bersch$^7$,
P.~Braun-Munzinger$^2$, W.~C.~Chang$^7$, W.~E.~Cleland$^6$,
M.~Clemen$^6$, J.D.~Cole$^4$, T.~M.~Cormier$^{9}$, Y.~Dai$^5$,
G.~David$^1$,
J.~Dee$^7$, O.~Dietzsch$^8$, M.W.~Drigert$^4$, K. Filimonov$^3$,
J.~R.~Hall$^9$, T.~K.~Hemmick$^7$, N.~Herrmann$^2$, B.~Hong$^2$,
C.~L.~Jiang$^7$, S.C.~Johnson$^7$, Y.~Kwon$^7$, R.~Lacasse$^5$,
Q.~Li$^{9}$, T.~W.~Ludlam$^1$, S.~McCorkle$^1$,
S.~K.~Mark$^5$, R.~Matheus$^{9}$, D.~Mi\'skowiec$^2$, 
E.~O'Brien$^1$, S.~Panitkin$^7$,
T.~Piazza$^7$, M.~Pollack$^7$, C.~Pruneau$^9$, Y.~J.~Qi$^5$,
M.~N.~Rao$^7$, E.~L.~Reber$^4$, M.~Rosati$^5$, N.~C.~daSilva$^8$,
S.~Sedykh$^2$, J.~Sheen$^9$, U.~Sonnadara$^6$,
J.~Stachel$^3$, H.~Takai$^1$, E.~M.~Takagui$^8$, S.~Voloshin$^6$,
T.~Vongpaseuth$^7$, J.~P.~Wessels$^3$, C.~L.~Woody$^1$,
N.~Xu$^7$, Y.~Zhang$^7$, Z.~Zhang$^6$, C.~Zou$^7$\\ (E877
Collaboration)} \address{$^1$ Brookhaven National Laboratory, Upton,
NY 11973\\ $^2$ Gesellschaft f\"ur Schwerionenforschung, Darmstadt,
Germany\\ $^3$ Physikalisches Institut, Universit\"at Heidelberg,
Heidelberg, Germany \\ $^4$ Idaho National Engineering Laboratory,
Idaho Falls, ID 
83415\\ $^5$ McGill University, Montreal, Canada\\ $^6$ University of
Pittsburgh, Pittsburgh, PA 15260\\ $^7$ The University at Stony Brook,
Stony Brook, NY 11794\\
$^8$ University of S\~ao Paulo, Brazil\\ $^9$ Wayne State University,
Detroit, MI 48202\\}

\maketitle

\begin{abstract} 
Inclusive double differential multiplicities of deuterons, $^3$H,
$^3$He, and $^4$He measured by E877 for 11.5 A GeV/c Au+Au
collisions at the AGS are presented.  
Light fragments at
beam-rapidity are measured for the first time at AGS energies.
Beam rapidity deuteron and $^4$He yields and transverse
slope parameters are found to be strongly dependent on impact parameter
and the shape of the
deuteron spectra is not consistent with that expected for
a simple thermal distribution.
The deuteron yields relative to proton yields are analyzed
in terms of a simple coalescence model.  While results indicate
an increase in source size compared to collisions of lighter
systems at the same energy, they are inconsistent with a
simple coalescence model reflected by a rapidity 
dependence of the coalescence parameter $B_d$.  
A new approach utilizing an expanding thermalized
source combined with a coalescence code is developed for studying
deuteron
formation in heavy-ion collisions.  The strong dependence of
deuteron yields on collective motion implies that
deuteron yields relative to those of protons can be used for
constraining source parameters.
\end{abstract}

\pacs{PACSnumber: 25.75.+r} 

\section*{Introduction}

The study of heavy-ion collisions at ultra-relativistic energies has
been pursued with the intent of observing matter at extreme
temperatures and densities, the eventual goal being the production and
study of
deconfined and/or chirally restored
matter.  Although the relativistic
heavy-ion program has
developed over a number of years, and various hadronic observables
have been studied, a complete characterization of the state formed
during the high density phase of the collision is still outstanding.

Measurements of light nuclei produced in the participant
region of heavy-ion collisions are of interest since model
calculations can be used to deduce measures of the source volume
from the yield of nuclear fragments
relative to that of protons \cite{sato,schwarzschild,mrow,mrow2,814}.
The yield of light nuclei is closely related to the two particle
correlation between nucleons induced by final state interactions
\cite{mrow2}. This can be incorporated into a theoretical framework 
from which one can deduce the density of the interaction region at freeze-out.

In this publication we will report on investigations of the
characteristics of spectra of light nuclei utilizing the E877
spectrometer, located at Brookhaven
National Laboratory's (BNL) Alternating Gradient Synchroton (AGS).
Spectra and yields of light nuclei, including
deuterons, $^3$H, $^3$He, and $^4$He will be presented and compared to
similar measurements at other beam energies and for various colliding
systems. 

\section*{Experimental Apparatus}

\begin{figure}
\centerline{\psfig{figure=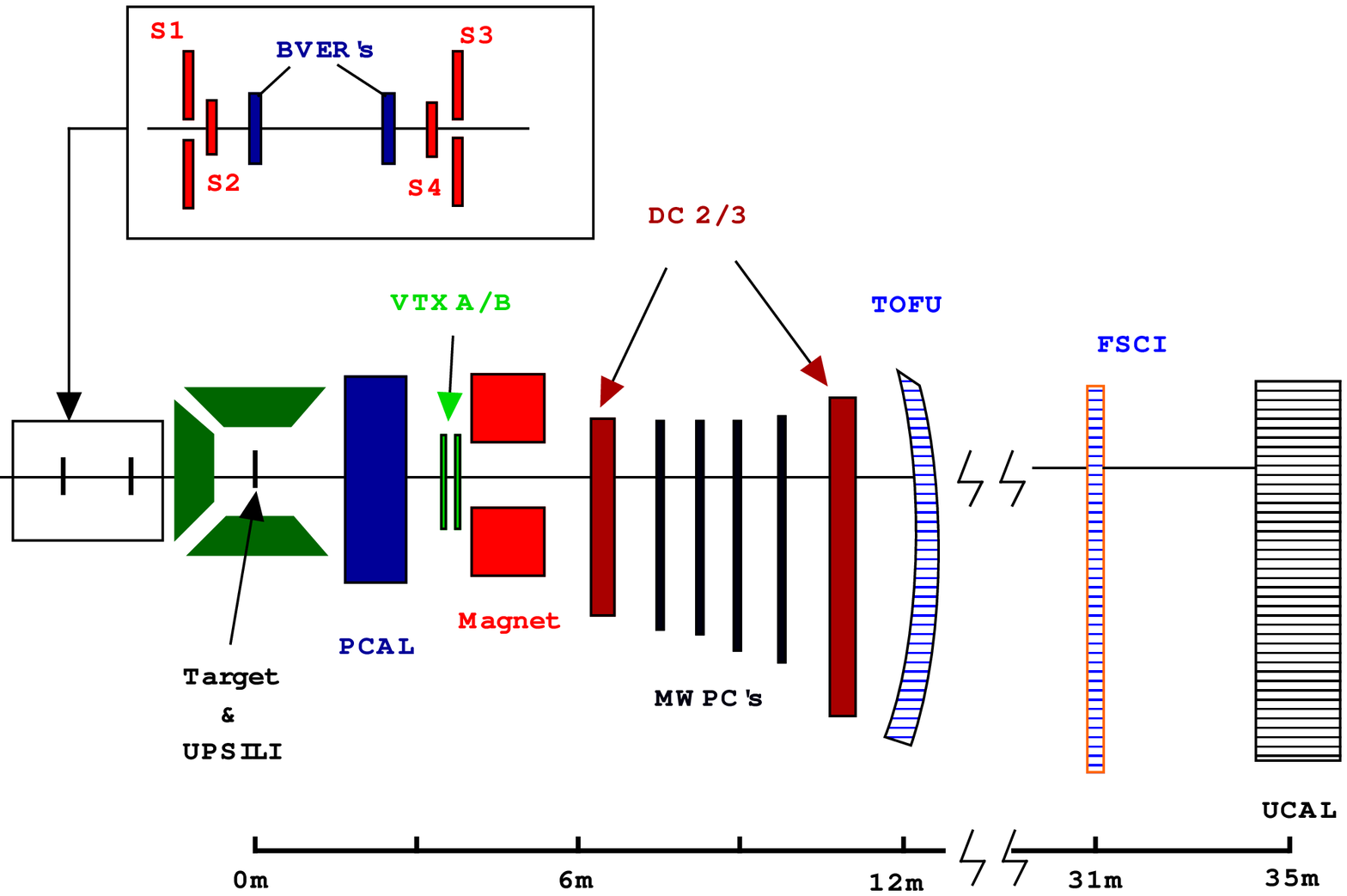,angle=90,height=17cm}}
\caption{E877 setup in the 1994 run}
\label{fig:apparatus}
\end{figure}

The E877 apparatus has already been described in \cite{814}.
Global observables, single particle spectra of protons,
pions\cite{piazza} and kaons \cite{877_qm97}
and two particle correlations \cite{panitkin,dariusz} 
as well as collective
flow \cite{877_flow} have been studied by E877 at the AGS.
The experimental setup
primarily consists of three components:  (1) a set of scintillators
and silicon detectors for beam definition; (2) a  near  $4\pi$
calorimetric coverage and a large acceptance charged particle
multiplicity measurement, providing global event properties; and (3) a
forward spectrometer, allowing measurements of identified charged particles.
Figure \ref{fig:apparatus} shows schematically the E877
experimental setup for the 1994 run.  The $z$ direction is defined to
be along the beam and the $y$ axis is defined to be out of the page
(vertical).


The data presented here were collected in the fall of 1994.  After
event selection cuts were applied, approximately $5.4 \cdot 10^6$
events for the 4\% highest $E_t$ collisions were available for analysis.

\subsection*{ Beam Definition }

The beam in the C5 beam line at the AGS for the 1994 run had
approximately $10^4$ particles per spill with a 1s spill length.
To verify the beam composition and direction, and provide the start
time of the experiment a number of detectors were placed upstream of
the target.

A series of four plastic scintillators (S1, S2, S3, S4) were used
for the time reference definition in the experiment.  
Scintillators S1 and S3 were ring-shaped
veto detectors while scintillators S2 and S4
were ellipsoidal detectors placed at $45^{\circ}$ relative to the
beam.  Fast rise time
phototubes were utilized on S2 and S4 providing the interaction time with a
typical Gaussian standard deviation of $\sigma = 35-40$ ps.

Two silicon
microstrip detectors (BVER's) measured
the beam angle in the $xz$-plane and were located between S2 and S4.
The detectors were 300 $\mu$m thick and composed of 320 strips with a
50 $\mu$m pitch.  This provided a measurement of the beam angle with a
resolution of about 
$40$ $\mu$rad and determined the 
$x$ coordinate of the beam at the target to 300 $\mu$m.
The BVERs were also used to tag and reject those events in which two
or more particles traversed the experiment within the same 1 $\mu$s
time interval.

To eliminate beam components
with incorrect nuclear charge number, a silicon semiconductor counter
(SILI) 87 microns thick was placed $2.5$ cm
upstream of the target to measure the energy loss of the beam
particles.  This detector provided a unit charge resolution of the
incoming beam particles.

\subsection*{Event Characterization}

For the 1994 run we used a AU target having a thickness corresponding
to 1\% of an interaction length.
Two large, highly segmented calorimeters were used to quantify the
centrality and reaction plane of the collisions.  Together
these detectors cover nearly 4$\pi$ in the center-of-mass frame.

The Target Calorimeter (TCAL) \cite{814_et} consists of 832
NaI(Tl)
crystals surrounding the target with a polar angle coverage of $48^{\circ} <
\theta < 135^{\circ}$ ($-0.88 < \eta < 0.81$).  Each crystal has a depth of
13.8 cm, or about 5.4 radiation lengths (0.34 hadronic interaction
lengths).

The Participant Calorimeter (PCAL) \cite{814_et} is a finely segmented 
lead/iron/scintillator calorimeter consisting of 16 azimuthal, 8
radial, and 4 depth sections. It covers a polar angle region of 
$1^{\circ} < \theta < 47^{\circ}$ ($0.83 < \eta < 4.7$).  There is a
small opening in the
PCAL ($-136 < \theta_x < 16$ mrad; $-11 < \theta_y < 11$ mrad) with iron
wedges along the inside which define the acceptance of the forward
spectrometer.

Events are characterized by their centrality as determined by the
transverse energy measured by the calorimeters.

\begin{equation}
E_t = \sum_i e_i sin\theta_i
\end{equation}

\noindent
where $e_i$ is the energy deposited in one cell of the calorimeter and 
$ \theta_i $ is the polar angle of the cell with respect to the beam
axis; the sum is performed
over all PCAL cells.  As was noted in \cite{814_et}, the overall
systematic error associated with the measurement of $E_t$ was
determined to be less than 4\% for all collisions considered in this
analysis.

The transverse energy is then used as a measure of the centrality of
the collision by noting that

\begin{equation}
\frac{d\sigma}{dE_t} = \frac{s}{n_tB}\frac{dN}{dE_t}
\label{eq:sig}
\end{equation}

\noindent 
where $\sigma$ is the cross-section, $B$ is the incident
number of beam particles, $s$ is the trigger down-scale factor ($s \ge
1$), and $n_t$ is the number of target atoms per unit cross-sectional
area.  For the experiment we used a Au target with an areal density of 
540 mg/cm$^2$.  This corresponds to a 1\% interaction probability for
incident Au ions at an calculated inelastic cross section of
$\sigma_{in}$ = 6.1
b.  Central collisions of a certain percentage ($w$) of the inelastic
cross section were then selected by adjusting the lower limit
E$_t^{low}$ such that \cite{877_flow}:

\begin{equation}
w = \frac{\int_{E_t^{low}}^{\infty}
\frac{d\sigma}{dE_t'}dE_t'}{\sigma_{in}}.
\label{eq:sig2}
\end{equation}

\subsection*{Forward Spectrometer}

The momentum and time-of-flight of all charged particles
accepted into the forward region are measured using two drift 
chambers, four multiwire proportional chambers and two time of
flight hodoscope walls.  These detectors are
also shown schematically in Figure ~\ref{fig:apparatus}.

In addition, a set of multiwire proportional chambers was placed 2 m
from the interaction region; the two chambers were separated from each
other by 25cm (VTXA and VTXB).  They were not used in
this analysis.

The analyzing magnet, situated downstream of the vertex chambers and
spanning from $z=260$ cm to $z=350$ cm,
could be run in two polarities in order to study possible systematic
errors in the
momentum measurement and to optimize the acceptance
for positive and negative particles.
The magnet had a
maximum field of $3,353 \pm 4$ Gauss with a polarity that bent positive
particles in the negative $x$ direction and an integrated $B \cdot dL$ 
of .3487 Tm \cite{panitkin}.

Two drift chambers (DC2 and DC3), located at $z=700$ cm and $z=1150$
cm from the target, respectively, provided track {\it x} and {\it y}
information and were used to determine
the rigidity of each track;  each drift chamber was composed of six
planes of wires which measured the position in the bending plane of the
magnet, {\em i.e.} the {\it xz}-plane, with a resolution of $250$ $\mu$m
($350$ $\mu$m) for DC2 (DC3).  The anode wires in DC2 and DC3 were
parallel to the {\it y}-direction, and the cathode planes in both
chambers were segmented into chevron shaped pads.  The signals from
the pads provided y 
measurements of charged particles with a resolution of typically a few
percent of the pad size.  Each drift chamber had two segmentation size
regions, resulting in $y$-resolutions of 2.3 mm (15 mm) in DC2 and 4.3
mm (36 mm) in DC3 depending on the pad size.

The multi-wire proportional chambers (MWPC's)
with vertical anode wires space 5.08 mm apart were
situated between the two drift chambers.  The MWPC's were used to 
ease tracking in the high multiplicity environment of
Au+Au collisions.

An array of 160 plastic scintillators (TOFU) \cite{tofu_nim}
aligned vertically and placed behind DC3 was used to measure
time-of-flight with a
time resolution in conjunction with the beam
start counters of about $85$ ps.  It also provided an 
additional y-measurement with a
precision of about 1 cm in the y-direction.
The FSCI (Forward Scintillator) was a second hodoscope which
also measured the time-of-flight at a distance of 31 m from the target
with a resolution of 350 ps.
It was utilized in this analysis only
to study of TOFU systematics since it had a lower resolution and
smaller acceptance than the TOFU.

\section*{Data Analysis}

\begin{figure}
\centerline{\psfig{figure=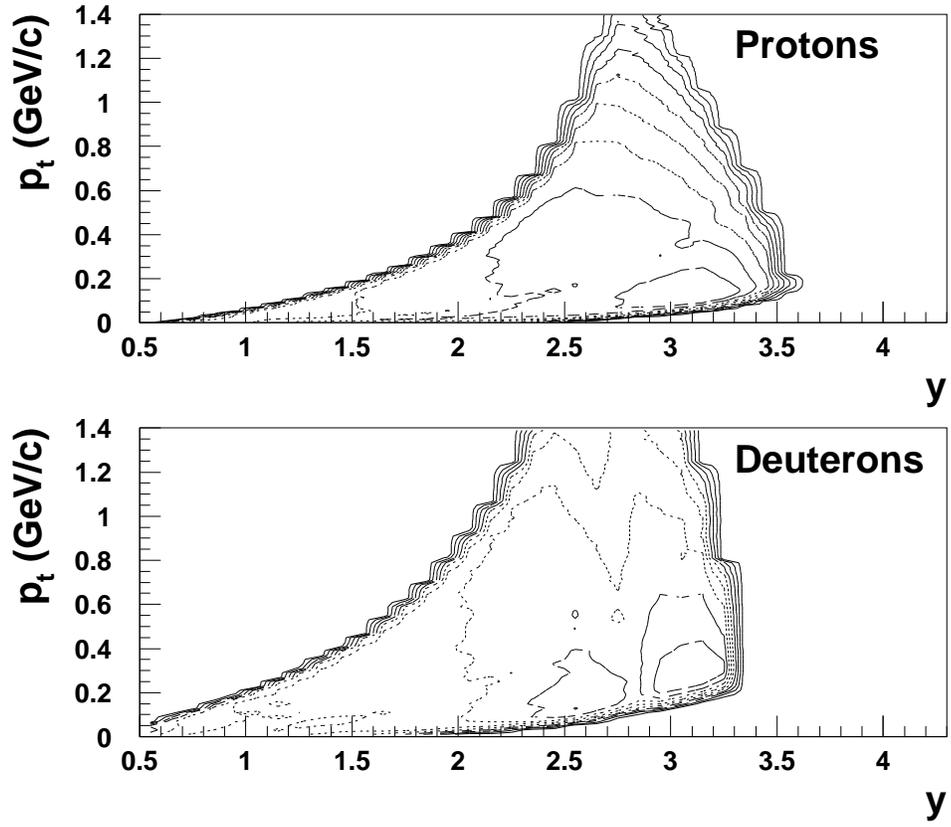,height=14cm}}
\caption{E877 acceptance in rapidity and transverse momentum for
protons and deuterons for the 1994 run.}
\label{fig:accept}
\end{figure}

Particle rigidity was determined from 
the radius of curvature of a track ($R$) through the magnet and the
value of the magnetic field $B$:

\begin{equation}
{\rm{rigidity}} = \frac{\sqrt{{p_x}^2+{p_z}^2}}{Ze} = RB
\end{equation}

\noindent
where $Ze$ is the charge of the particle.  The vertical momentum
component, $p_y$, is calculated from
the vertical position of the track at the TOFU.  The
reconstructed mass of each particle is then given by

\begin{equation}
m^2c^4 = \frac{p^2c^2}{\beta^2 \gamma^2}
\label{eqn:mass}
\end{equation}

\noindent
where $\beta$ and $\gamma$ are calculated from the time-of-flight
measurement coupled with the calculated path length to the associated
TOFU scintillator.
After all momentum calculations, Fig.~\ref{fig:accept} shows the E877
acceptance for protons and
deuterons as a function of transverse momentum and rapidity for the
1994 run.

\begin{figure}
\centerline{\psfig{figure=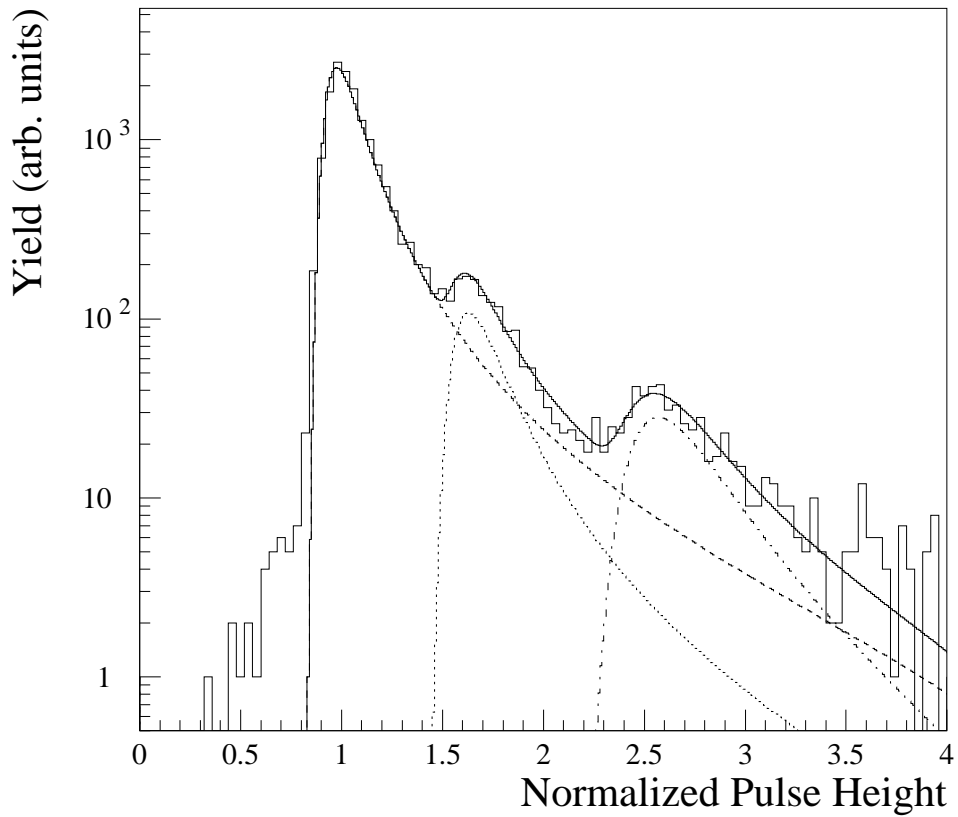,height=14cm}}
\caption{The TOFU pulse height distribution for all particles
  at beam rapidity and from 0.5 to 1.0 GeV/c in $p_t$.  The pulse
  height is normalized to the most probable energy loss of a single
$Z=1$ minimum
  ionizing particle.  The histogram
  is the measured TOFU distribution.  The solid curve is a full fit
  while the dashed and dotted curves correspond to the individual
  contributions from $Z = 1$, twice $Z = 1$, and $Z = 2$
  peaks.}
\label{fig:fit_he4}
\end{figure}

The charge of each particle is determined from the energy deposition
in the TOFU slat.  Shown in Figure~\ref{fig:fit_he4} is the distribution
of pulse-heights for those particles corresponding to the deuteron
mass peak at beam rapidity.  A clear $Z=1$ peak is visible as well as
a peak corresponding to $Z=2\cdot1$, from two $Z=1$ tracks
striking the
same TOFU slat.  In addition, the $Z=2$ peak corresponding to $^4$He
is clearly visible.
The resulting distribution is fit to three Landau functions with a
non-linear response due to the saturation of the photomultiplier tubes
associated with the TOFU.
The standard E877 analysis cut on the TOFU pulse-height distribution
at $1.4$ times the 1 MIP pulse height
corresponding to about a $10$\% loss in the $Z=1$
yield with no contamination from $Z=2$.

\subsection*{Deuterons}

Since $\delta m/m \propto \delta p/p$ and since the momentum
and time of flight measurements worsens at larger momenta, 
deuteron selection is complicated by the
tail of the proton peak at high momenta.
To account for this
effect, a systematic background subtraction technique
was developed\cite{johnson}.  In each rapidity and transverse momentum
bin the yield of
deuterons is calculated by fitting the mass distribution in this
region to the proton and deuteron mass peaks plus a background between 
these peaks due to misidentified protons from resonance decays.

\begin{figure}
\centerline{\psfig{figure=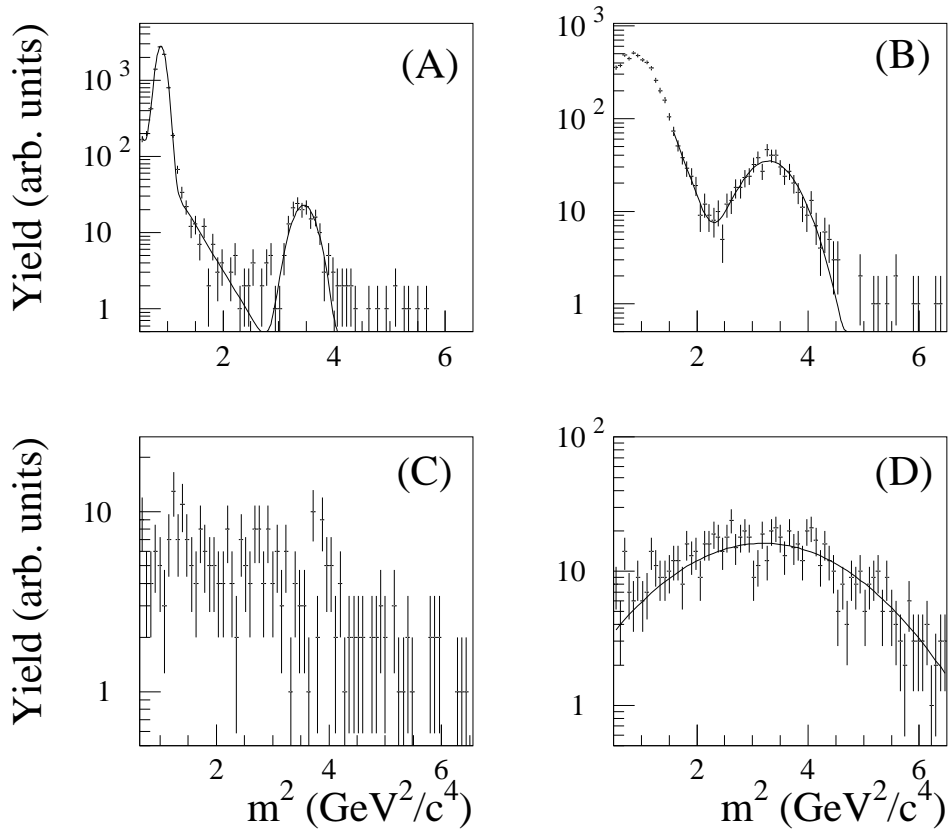,height=14cm}}
\caption{Fits to the mass-squared distribution
for several regions of momentum and rapidity space, where rapidity
is calculated assuming the deuteron mass: (A) $y=1.4-1.5$, $p_t=.1-.12$
  GeV/c fit with two Gaussians and an exponential background; (B)
  $y=2.2-2.3$, $p_t=.68-.70$ GeV/c fit with two Gaussians; (C)
  $y=2.8-2.9$, $p_t=.50-.52$ GeV/c; (D) $y=3.0-3.1$, $p_t=.20-.22$ fit
  with a single Gaussian.
  See
text for details.}
\label{fig:mult_par_fit}
\end{figure}

Shown in Figure \ref{fig:mult_par_fit} is the
reconstructed distribution of mass-squared in several
$p_t = \sqrt{({p_x}^2+{p_y}^2)}$ and $y_{deut} = 0.5 \ln(E+p_z)/(E-p_z)$
bins, where the rapidity was calculated assuming the deuteron mass.
At low momenta, the deuteron and proton peaks are
clearly distinguishable.
In this region of phase space, the
mass-squared distribution is described by two Gaussians and either an
exponential
or linear background\footnote{The systematic error associated with
  the choice of the background function has been found to be less than
  5\% .}.  At
higher momenta, the proton and deuteron mass-squared distributions broaden,
due to the momentum resolution of the apparatus, and the
background is minimal.  In this region of phase-space, shown
in Figure \ref{fig:mult_par_fit}B, the
mass-squared distribution is described by two Gaussian distributions.  In
the $2.6< y <3.0$ region, the
momentum resolution smearing coupled with the approximately equal
particle yields for protons and deuterons, causes the peaks to overlap
(Figure ~\ref{fig:mult_par_fit}C), resulting in a
loss of separation of deuterons from protons in this region.  However,
at deuteron rapidity $y>3.0$
the contamination of protons is
minimal because the proton would need to be traveling at twice the beam
momentum.  Thus, a
deuteron peak is again identifiable (Figure
\ref{fig:mult_par_fit}D) and the yield can be determined with a single
Gaussian.  The widths of the proton and deuteron mass peaks were found 
to be consistent with the known time of flight and momentum
resolution of the experiment.

The yield of deuterons ($N_d$) in each small bin of phase space is
determined from the parameters of a Gaussian fit to the deuteron
peak:

\begin{equation}
N_d = \sigma_d A_d\sqrt{2\pi}/\Delta m^2
\end{equation}

\noindent
where ${\sigma_d}^2$ and $A_d$ are the variance and amplitude of the
Gaussian distribution fit to the deuteron peak, and $\Delta m^2$ is
the bin width.  Those regions of momentum space that fall
between the first two regions outlined above, Figure
\ref{fig:mult_par_fit}A and B, respectively, have been fit with both
two Gaussians as well as two Gaussians plus a background in order to
determine the systematic error due to the choice of fit funcion.
The resulting yields of deuterons agree within 5\%.

\subsection*{Tritons}

\begin{figure}
\centerline{\psfig{figure=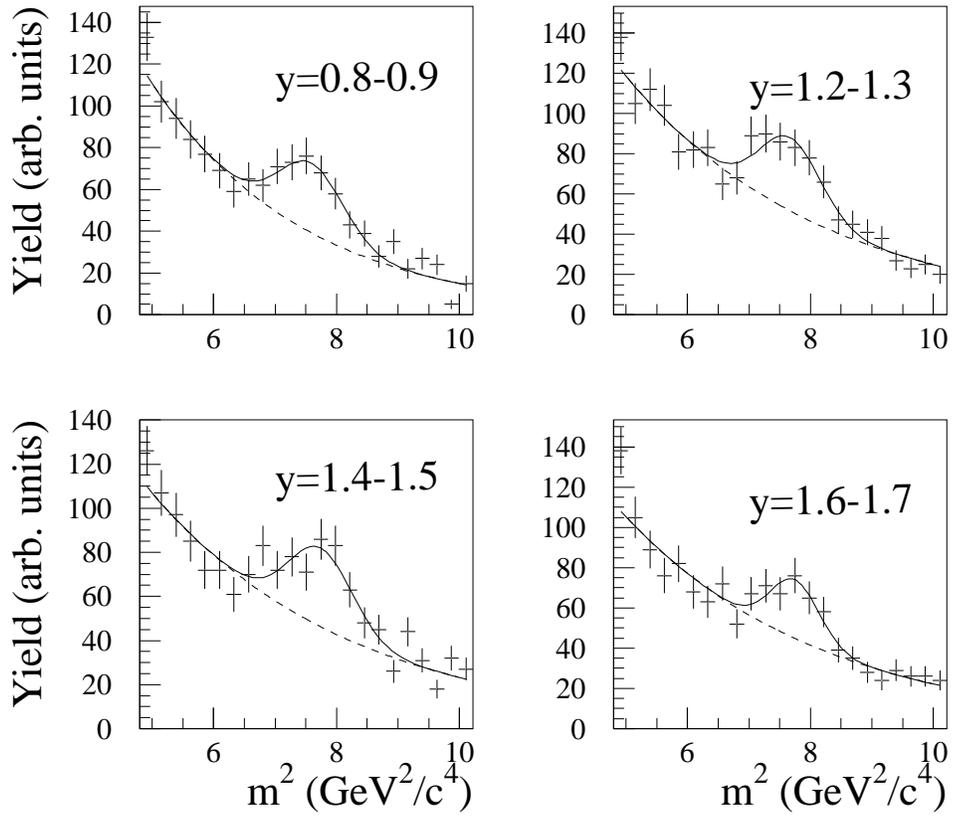,height=14cm}}
\caption{Mass-squared distributions for various rapidities (calculated assuming
  the triton mass) fit with a Gaussian at the mass of the triton plus an
  exponential background from the deuteron peak.}
\label{fig:tri_mass}
\end{figure}

The extraction of a triton signal is complicated by the fact that its
mass-squared distribution is wider than the deuteron's due to multiple
scattering
and its yield is 1-2 orders of magnitude 
lower than that of the deuteron.
However, a technique similar to that used for the deuterons can also
be utilized to measure the yield of tritons in the collisions studied
here.  In 
Figure \ref{fig:tri_mass} is shown the reconstructed mass-squared distribution
in various $y_{trit}$ bins for $p_t < 0.5$ GeV/c.  A clear triton peak
is visible.  The yield of tritons
was calculated for each rapidity and $p_t$ bin by fitting this
distribution with a Gaussian at the mass of the triton with an
exponential background from the deuteron tail;  the choice of an
exponential or Gaussian background was found to have less than a 5\%
effect on the resulting yields.  Several examples of these fits are also
shown in Figure \ref{fig:tri_mass}.

\subsection*{Helium Isotopes}

\begin{figure}
\centerline{\psfig{figure=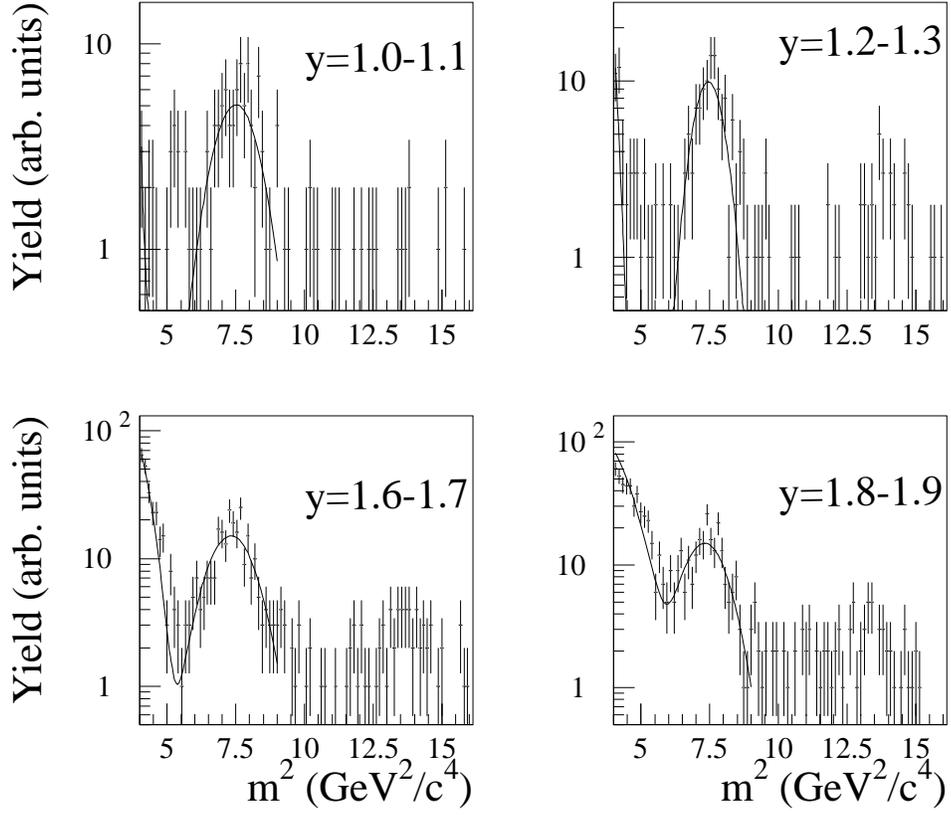,height=14cm}}
\caption{Mass-squared distributions for various
  rapidities (calculated assuming the $^3He$ mass) and $p_t < .5$
  GeV/c.  A clear peak is seen at the $^3He$ mass which is lost under
  the proton tail for rapidities above $y \sim 2.0$.  The resulting
distribution is fit with two Gaussians at the proton and $^3$He mass peaks.}
\label{fig:he3_mass}
\end{figure}

Helium isotopes can be separated from
$Z=1$ species by cutting on energy deposition in the
scintillators as described earlier.  $^3$He are identified by
recalculating the mass-squared according to
eqn.~\ref{eqn:mass} assuming $Z=2$.  This will cause all $Z=1$
particles which pass the scintillator pulse-height cut to be found at
double their actual mass.  The
$^3$He peak in this case is found between, and clearly differentiated
from, the proton and deuteron
residual peaks, respectively located at $m^2 \approx 4$ and $m^2
\approx 14$ (GeV/c)$^2$, as shown in Figure \ref{fig:he3_mass}.
The $^3$He peak is
buried under the proton tail for $y > 2.0$ and, in the other rapidity
bins, the yield of $^3$He 
may be extracted by fitting the mass-squared distribution
by two Gaussians corresponding to the $^3$He and proton peaks.

The method described above, {\em i.e.} 
cutting on the energy loss in the TOFU and, subsequently, fitting the
peak in the $m^2/Z^2$ spectrum, was used to determine the deuteron,
triton and $^3$He yield.  However, this method cannot be used to
determine the yield of $^4$He because the $^4$He peak in $m^2/Z^2$
space is buried under the
significantly larger deuteron peak.  We therefore use a complimentary
approach to identify $^4$He:  we first cut on the $m^2/Z^2$
peak and then fit the $Z=2$ peak in the TOFU pulse height distribution.
As described earlier, this distribution has two peaks, corresponding
to one and two $Z=1$ particles, at all but beam rapidity where an
additional 
$Z=2$ peak is evident as shown in Figure \ref{fig:fit_he4}.  These
peaks are fit, at beam rapidity for $p_t<2$ GeV/c, to Landau
distributions and the yield is then determined
from the integral under the fit to the $Z=2$ distribution.
The overall systematic errors from this procedure is estimated to be
less than 10\%.

\section*{Results}

\subsection*{Mid-rapidity}

\begin{figure}
\centerline{\psfig{figure=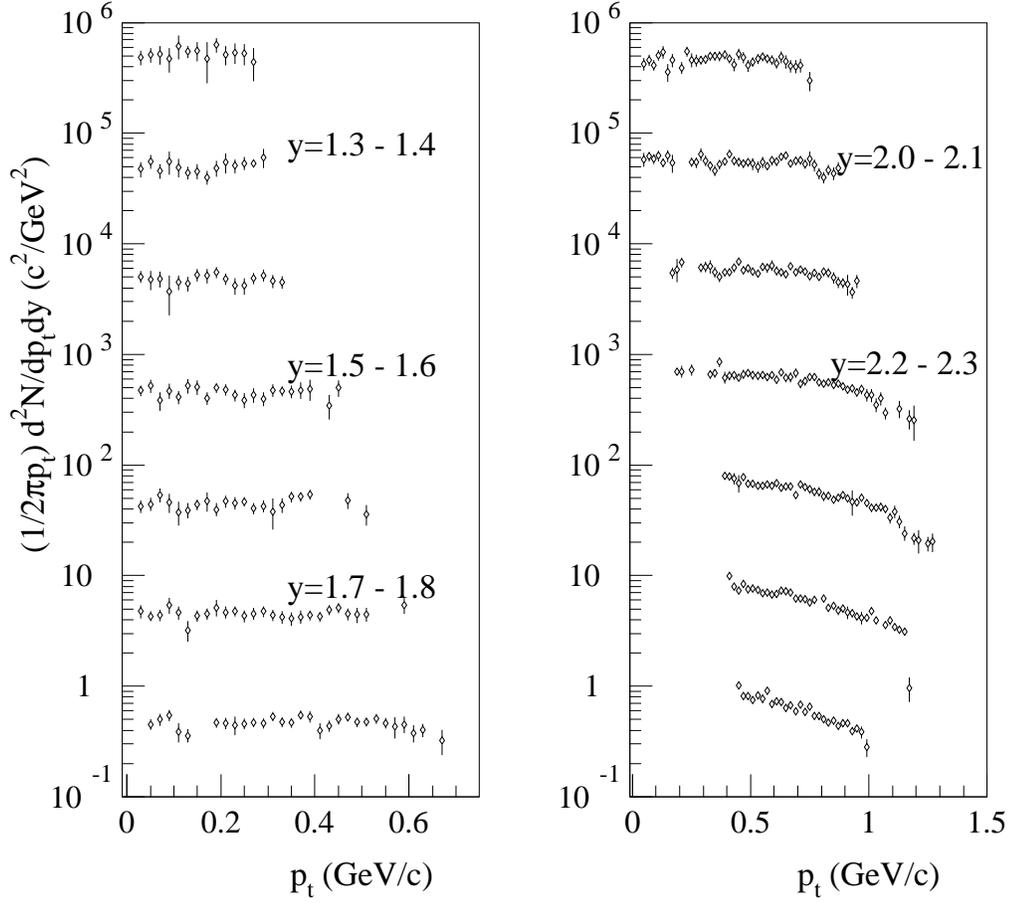,height=15cm}}
\caption{Invariant multiplicities of deuterons for the 4\%
highest $E_t$ events.  The spectra are binned in units of .1 in rapidity
and 20 MeV in $p_t$. The $y=2.5-2.6$ and
$y=1.8-1.9$ spectra are properly normalized and the yields for the lower
rapidity bins are multiplied by successive powers of 10 for clarity of
the presentation. {\em e.g.} The $y=1.7-1.8$ bin is multiplied by a
factor of 10, the $y=1.6-1.7$ bin is multiplied by a factor of 100,
etc.  The error bars reflect the statistical uncertainties.} 
\label{fig:deut_spec}
\end{figure}

Shown in Fig.~\ref{fig:deut_spec} are deuteron invariant
multiplicities for the 4\% highest $E_t$ collisions as a
function of $p_t$ and $y$ with bin widths of 20 MeV in $p_t$ and 0.1
in $y$.
The invariant multiplicity is flat as a function
of $p_t$ and $y$ over most of the measured range.

\begin{figure}
\centerline{\psfig{figure=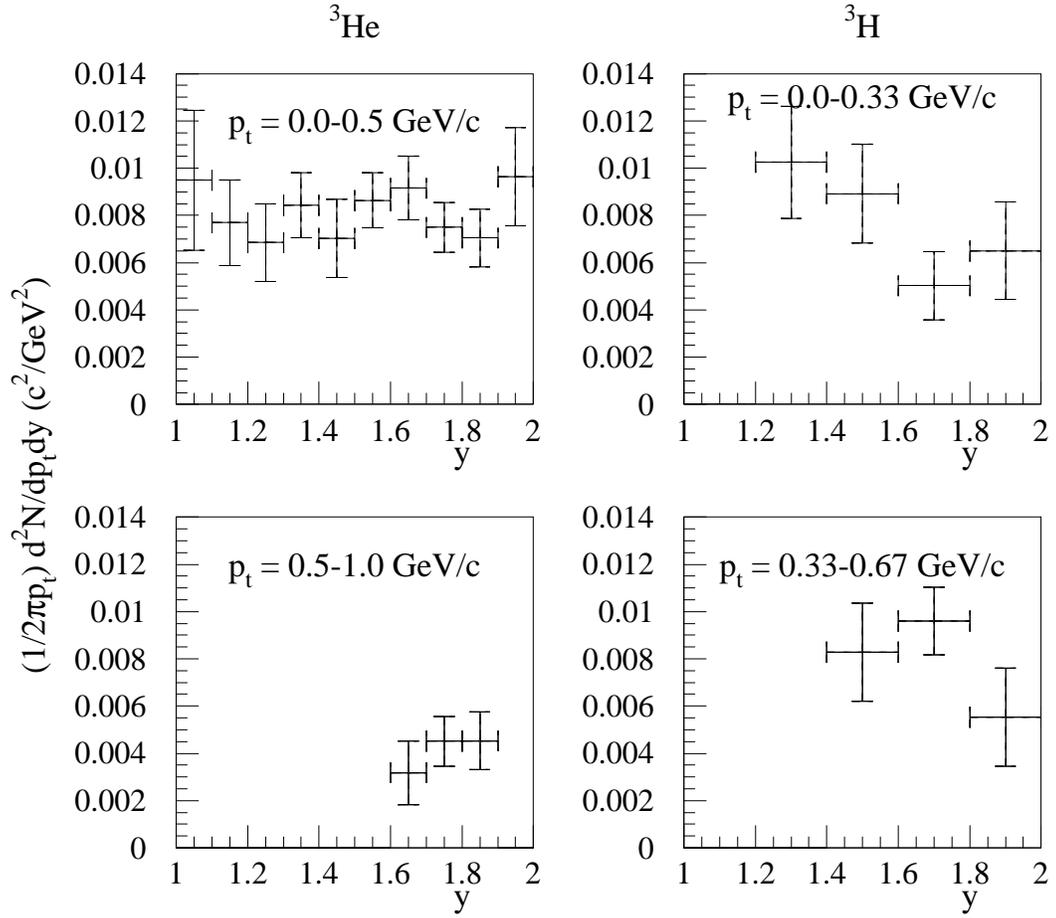,height=15cm}}
\caption{Invariant multiplicities of $^3$He and $^3$H for the 4\% inclusive
$E_t$ bin.  Plotted is the invariant cross section at the center
of the bin assuming an underlying thermal distribution similar to what 
is measured.  Error bars reflect statistical uncertainties only.}
\label{fig:h3_he3}
\end{figure}

In Figure \ref{fig:h3_he3} are shown the measured invariant
multiplicities of $^3$H and 
$^3$He for the same centrality.
Plotted is the differential cross
section at the center of the bin assuming an underlying thermal
distribution with a shape similar to what is measured \footnote{The
resulting slope and magnitude of the $p_t$ spectra was found to be
insensitive to a wide range or reasonable assumptions of the
true distribution in the acceptance correction.}.
The resulting
measured cross section is not affected significantly by the assumed
distribution.
The distributions
are presented as a function of rapidity for two $p_t$ regions.
The error bars reflect the statistical uncertainty in the fit
procedure and
do not include the systematic errors, estimated to be about $20\%$ for
both $^3$He and $^3$H, dominated by the uncertainty in the
assumed background.  

\subsection*{Beam-rapidity}

\begin{figure}
\centerline{\psfig{figure=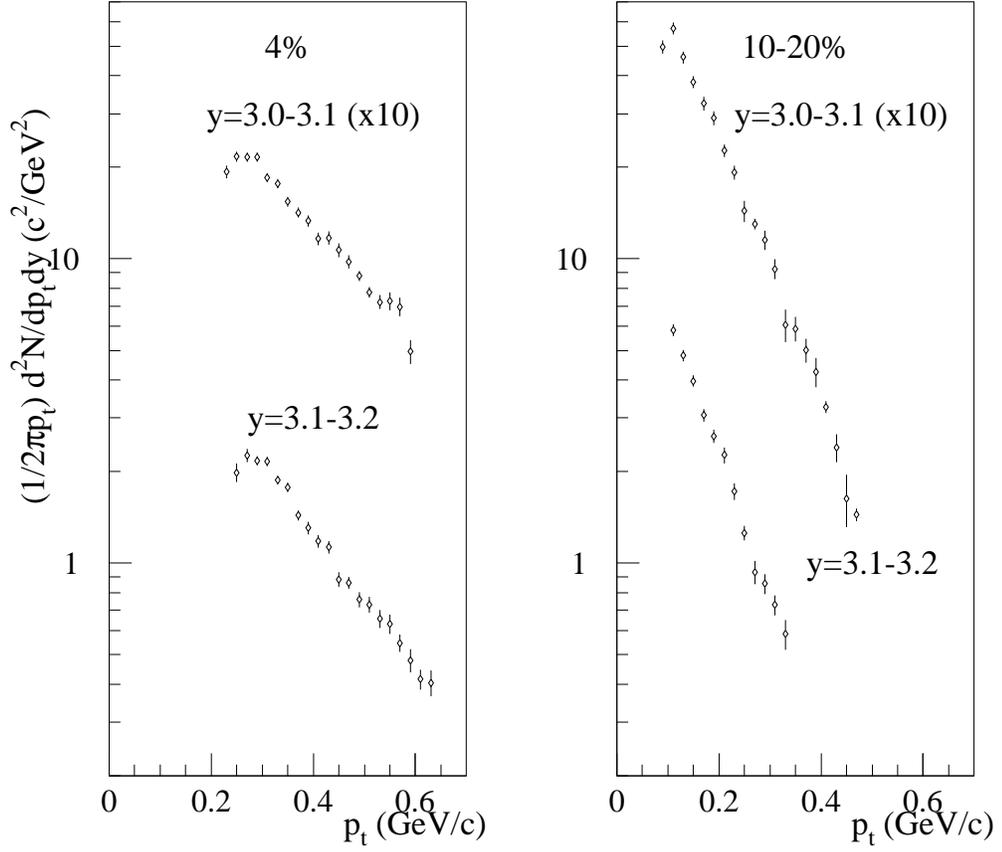,height=15cm}}
\caption{Invariant multiplicities of deuterons near beam rapidity for the
collisions with $E_t$ in the highest 4\% and the  10-20\% range.  Plotted
is the invariant cross section at the center
of the bin assuming an underlying thermal distribution similar to what 
is measured. Error bars reflect statistical uncertainties only.}
\label{fig:beam_letter}
\end{figure}

In Figure \ref{fig:beam_letter} are shown invariant multiplicities for
deuterons at beam rapidity ($y_{beam}=3.2$) for the 4\% and 10-20\%
highest $E_t$
events.  As noted above, deuterons at beam rapidity are easily
extracted from the mass-squared distribution.
The transverse momentum distribution of beam rapidity deuterons
for the 4\% highest $E_t$ are distinctly
harder, {\em i.e.} show a larger inverse slope parameter, than
deuterons from the lower $E_t$ collisions.  Over the range in
$p_t$ in which we measure, a fit to an $m_t$ 
Boltzmann distribution in the 3.0-3.1 rapidity bin gives $T_B = 49.6 \pm
1.4$ MeV and $T_B = 83 \pm 2$ MeV for the 4\% and 10-20\% $E_t$
bins, respectively.
However, the reduced $\chi^2$ of each of these fits is greater than 4,
implying a Boltzmann shape does not describe
the measured distributions well.

\begin{figure}
\centerline{\psfig{figure=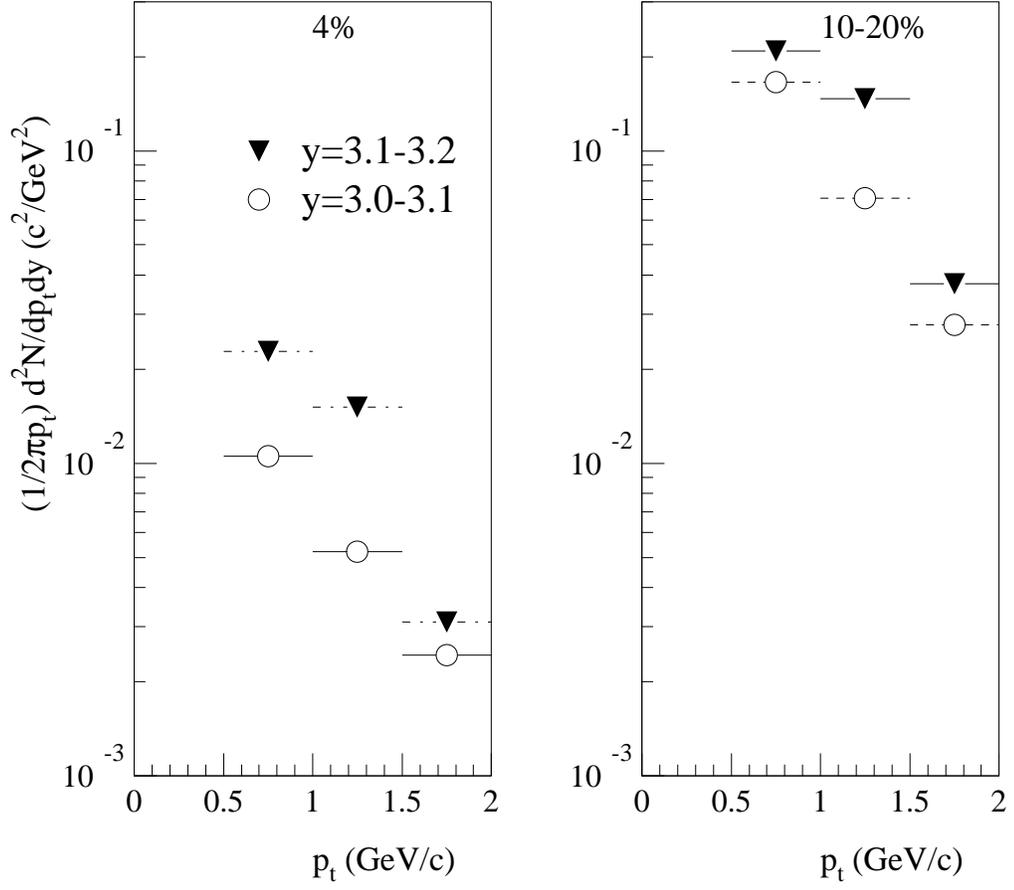,height=15cm}}
\caption{Beam rapidity $^4$He invariant multiplicities as a function of transverse
momentum for the 4\% and 10-20\% $E_t$ bins. Error bars reflect
statistical uncertainties only.}
\label{fig:he4_letter}
\end{figure}

In beam rapidity $^4$He spectra are shown in Figure \ref{fig:he4_letter}.
In the lower
$E_t$ range (10-20\%), the $^4$He yield is a factor of 10 higher
than for the higher $E_t$ (4\%), indicating that beam rapidity
$^4$He are primarily produced by projectile fragmentation.

\section*{Model Comparisons}

Light nuclei from heavy-ion collisions have historically been
studied via a coalescence model \cite{schwarzschild,butler} which
assumes cluster formation takes place at freeze-out.  Since the
binding energy of
the deuteron is $2.24$ MeV, it is easily broken apart in the
fireball region where current models estimate the temperature to
be 120-150 MeV \cite{thermal,thermal2}.
In the coalescence
formulation \cite{butler} the probability of
forming a deuteron is greatest when a proton and neutron at freeze-out
have a small relative momentum.
Assuming that the proton and neutron distributions are similar, the deuteron 
momentum distribution is then given by

\begin{equation}
E_d \frac{d^3N_d}{dp^3}(p_d) = B_d \left(E_p
\frac{d^3N_p}{dp^3}(p_p) \right)^2
\label{eqn:coal_1}
\end{equation}

\noindent
where the particle momenta obey $p_d = 2p_p$ which can be easily
generalized to heavier fragments with mass number $A$ as 

\begin{equation}
E_A \frac{d^3N_A}{dp^3}(p_A) = B_A \left( E_p \frac{d^3N_p}{dp^3}(p_p)
\right)^A
\label{eqn:coal_2}
\end{equation}

\noindent
where $p_A = A p_p$.  In eqns.~\ref{eqn:coal_1} and \ref{eqn:coal_2},
$B_A$ ($B_d$ in the case of the deuteron) is a phenomenological
parameter known as the coalescence constant.  In proton-nucleus
collisions $B_A$ has been related to
the interplay between the binding energy of the deuteron and the
optical potential of the target nucleus \cite{butler}. For
nucleus-nucleus collisions it was recognized that the idea of a
nuclear optical potential was no longer meaningful and Schwarzschild and
Zupan\v{c}i\v{c} \cite{schwarzschild} expressed the
coalescence parameter in terms of a momentum difference $p_0$.
A proton and neutron would coalesce to form a deuteron if, at
freeze-out, their momenta differ by less than $p_0$.  More
generally, for the case of a source of Z protons and N neutrons with
mass number of the cluster A,

\begin{equation}
B_A = \left( \frac{2s_A + 1}{2^A}\right) \frac{{R_{np}}^N}{N!Z!}
\left( \frac{4\pi}{3} {p_0}^3 \right)^{A-1}
\end{equation}

\noindent
where $s_A$ is the spin of the cluster and $R_{np}$ is the neutron to
proton ratio in the source.
Therefore, the coalescence parameter in these models is simply related
to the momentum difference between corresponding protons and neutrons
and is described by a simple step function in the coalescence
probability at $p_0$.  Note that no dependence on the colliding system
exists in this model which described deuteron and $^3$H/$^3$He
production very well in
nucleus-nucleus reactions at Bevalac \cite{nagamiya,wang} energies
(about $0.1-2$GeV$\cdot$A) and in
p-nucleus collisions, at FNAL \cite{cronin}
energies.  For all of these energies, only a single constant $B_A$ was
needed to describe the light fragment distributions based on proton
spectra in minimum bias collisions. 
However, for nucleus-nucleus collisions at AGS energies and above
a dramatic drop below the Bevalac values in the coalescence parameter
and a strong dependence on the centrality of the collision
\cite{814,na44pa} was observed.
For beam energies where nuclear collisions produce a large
number of secondary particles, {\em i.e.} AGS energies and above,
$B_d$ was no longer independent of beam energy or composition
presumably because the source decoupled at a size larger than the
mean size of a deuteron, $R_{rms} = 2.1$ fm \cite{nagle}.
In this case, $B_d$ was theoretically deduced to be inversely
proportional to the volume of the source at freeze-out.  In the model
of Sato and Yazaki \cite{sato}, for example,

\begin{equation}
B_d = \frac{3}{4} \frac{(8\pi)^{3/2}}{Z!N!} \left[ \frac{\nu_d \nu}
{(\nu_d+\nu)} \right]^{3/2}
\end{equation}

\noindent
where $\nu_d=.20fm^{-2}$ is the Gaussian wave function parameter
for a deuteron and $\nu$ is directly related to the radius of the
system at freeze-out by

\begin{equation}
R_{rms} = \sqrt{{R_x}^2+{R_y}^2+{R_z}^2} = (3/2\nu)^{1/2}
\label{eqn:rms}
\end{equation}

\noindent
where $R_x$, $R_y$ and $R_z$ are the Gaussian radii.
Other models \cite{mrow} also utilize a wave-function description and
a density matrix formalism to describe deuteron formation.  With the
exception of large
cascade codes \cite{nagle,kahana} as well as recent hydrodynamically
motivated source parametrizations \cite{heinz}, however, none of them
explictely
account for transverse collective expansion.  Though such an approach is not
excluded from the Sato and Yazaki approach \cite{sato}, nothing beyond 
sources with no
space-momentum correlations were considered in their
presentation. Further, though there have been some attempts 
at a fully relativistic treatment of coalescence \cite{dover}, this
generalization is non-trivial because of difficulties in
the proper relativistic treatment of bound states \cite{mrow}.

\subsection*{Coalescence Parameter $B_d$}

The coalescence parameter, as defined in Eq.~\ref{eqn:coal_1}, can be
calculated from E877 measurements of
protons \cite{piazza} and deuterons.  To properly calculate $B_d$
for our data set, the fraction of
protons from $\Lambda$ decays must be subtracted from the measured
proton spectra in order to obtain the proton distribution at
freeze-out.  This correction was performed utilizing the measured
$\Lambda$ yield in Au+Au collisions \cite{e891} and knowledge of the
acceptance and reconstruction efficiency of the experimental
apparatus.  Over the measured acceptance region, the hyperon
contribution to the proton spectra was found to be no more than 10\%.

\begin{figure}
\centerline{\psfig{figure=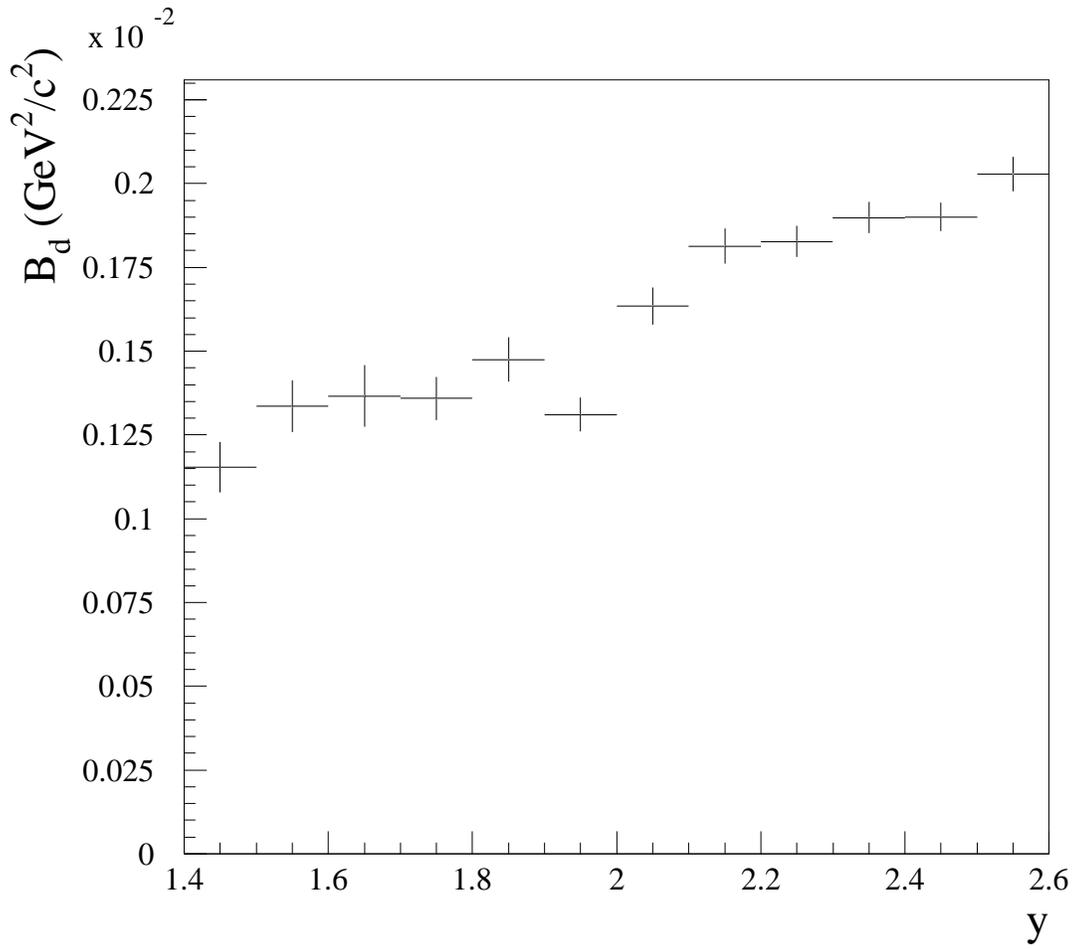,height=15cm}}
\caption{Coalescence parameter, $B_d$, as a function of rapidity for
the 4\% highest $E_t$ events. Error bars reflect statistical
uncertainties only.}
\label{fig:b2_y}
\end{figure}

After correcting for hyperon contributions, $B_d$ can be calculated as a 
function of $p_t$ and y.  We notice, within the acceptance of the
present measurement, no variation of $B_d$ as a function of
$p_t$.  However, there is a significant rapidity 
dependence of the average $B_d$ for all measured $p_t$ as shown in
Figure \ref{fig:b2_y}.
If $B_d$ is related to volume at freeze-out, such a
result may imply a changing effective freeze-out volume as a function of
rapidity.  The measured change in $B_d$ of nearly a factor of two from
mid-rapidity to $y=2.6$ would correspond to a decrease in the
effective source radius of $1.25$.

\begin{figure}
\centerline{\psfig{figure=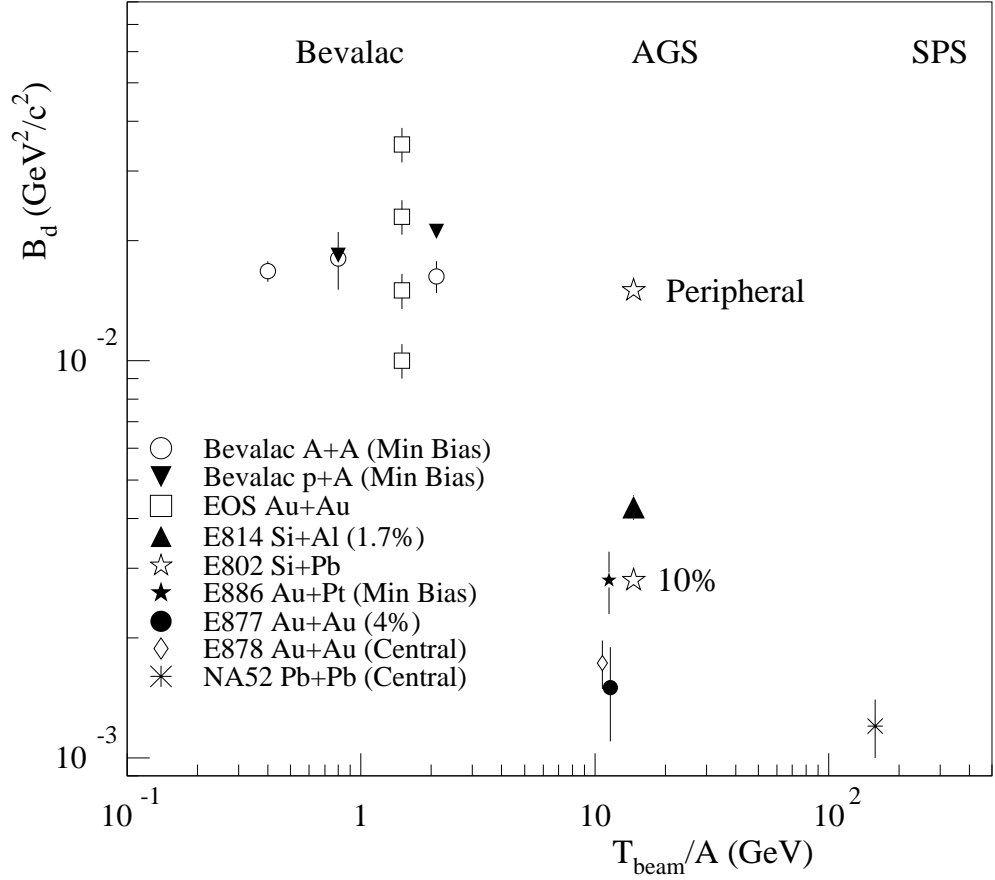,height=15cm}}
\caption{$B_d$ for a variety of systems and energies.  The values in
parenthesis are the centrality cut, where `Min Bias' corresponds to minimum
bias events. The Bevalac A+A data are for a variety of systems.
Since no dependence on colliding system was found,
the average value is shown here.  The recent EOS data are measured in
four separate multiplicity bins with the highest multiplicity
corresponding to the lowest value of $B_d$. The data are from 
[5,18,19,28-31].}
\label{fig:values}
\end{figure}

Figure~\ref{fig:values} shows a comparison of the coalescence parameter
deduced from experiments at Bevalac \cite{nagamiya,wang} energies with 
those obtained at the AGS in
in Si + Al \cite{814}, Si + Pb \cite{802}, Au + Pt \cite{886}, and Au
+ Au \cite{e878} collisions.  The
error on the Au+Au data point corresponds to the minimum and maximum
range of $B_d$ values shown in Figure~\ref{fig:b2_y} including the
statistical error bars.
All presently measured values of $B_d$ for high $E_t$ Au+Au collisions are
significantly lower than the results for Si+Al and Si+Pb.  This is
consistent with the interpretation
that $B_d$ depends on the volume at freeze-out.  Two proton and two
pion correlations measurements have found that the effective
nucleon source size is larger for the Au+Au system than for lighter systems
\cite{panitkin,dariusz}.

The coalescence parameter measured in these collisions
is nearly one-tenth of that measured in central Au+Au
collisions at Bevalac,
implying in the models of \cite{sato,mekjian} that the volume at
feeze-out in AGS Au+Au collisions is five to ten times that at Bevalac
energies.  Such a simple geometrical interpretation would be
inconsistent with two particle
correlations measured at the AGS \cite{dariusz,panitkin}.

A possible reason for such an inconsistency is that the models which
deduce an inverse volume dependence \cite{sato,mekjian} do not account
for collective motions in the source.  As has been shown in
\cite{mattiello}, the yield of deuterons is highly dependent on the
amount of collective motion in the source. 
Hence, $B_d$ is not simply related to the volume
but also the position-momentum correlations at freeze-out.  If the
source is perfectly correlated, then proton-neutron pairs that
freeze-out near one another in space will have similar momentum
vectors.  The deuteron yield
would be larger in this case than for a random source.
As such, any consistent model of deuteron production must account for
correlations in the source.

Therefore, in order to completely describe deuteron production, a more
sophisticated model of the source that includes dynamical correlations
must be formulated.  In large cascade codes it is difficult to study
the influence of the source parameters.  We thus used a simple
dynamical model in the following section to study such effects.

\begin{figure}
\centerline{\psfig{figure=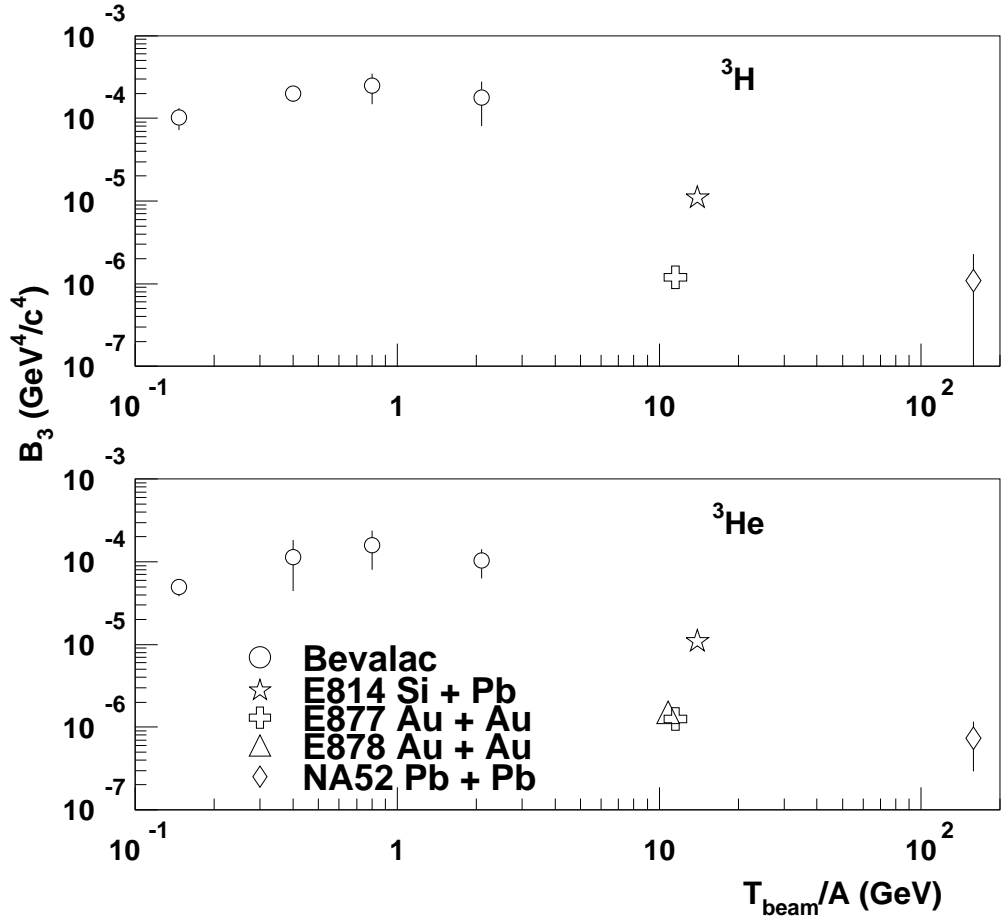,height=15cm}}
\caption{$B_3$ for a variety of systems and energies.
The Bevalac A+A data are for a variety of systems.
Since no dependence on colliding system was found,
the average value is shown here.  AGS and SPS values are only for
central events while Bevalac values are minimum bias.
The data are from 
[5,18,19,28-31].}
\label{fig:b3}
\end{figure}

In Fig.~\ref{fig:b3} is the corresponding coalescence parameter for
$^3$H and $^3$He extracted from this data set compared with Bevalac
\cite{nagamiya}, AGS \cite{814,e878}, and SPS \cite{na52} values.
Similar to what is observed in $B_d$, the three nucleon coalescence
parameter drops dramatically at AGS energies, consistent with an
increase in the effective source size.

\section*{A Dynamical Model for Deuteron Formation}

An analytic, hydrodynamically motivated model,
devised by Chapman, {\em et al.} \cite{chapman} has been
implemented to describe the particle source distribution in heavy ion
collisions \cite{wiedemann}.  This
model uses an emission function ($S$) for a finite, expanding, locally
thermalized source

\begin{equation}
E_K \frac{d^3N}{d^3K} = \int d^4x S(x,K)
\end{equation}

\noindent
where $x$ and $K$ are the position and momentum 4-vectors, respectively.
The source includes longitudinal and transverse collective motion with
an underlying temperature, given by

\begin{equation}
S(x,K) = \frac{M_t \cosh(\eta - Y)}{(2\pi)^{7/2}\Delta \tau}
exp \left[-\frac{K\cdot u(x)}{T} - \frac{(\tau-\tau_0)^2}{2(\Delta
\tau)^2} - \frac{r^2}{2R^2} - \frac{(\eta - \eta_0)^2}{2(\Delta
\eta)^2} \right]
\label{eqn:source}
\end{equation}

\noindent
where

\begin{eqnarray*}
\eta & = & \frac{1}{2} ln[(t+z)/(t-z)] \\
\tau & = & \sqrt{(t^2-z^2)} \\
Y & = & \frac{1}{2} ln[(1+\beta_l)/(1-\beta_l)] \\
K \cdot u(x) & = & M_t\cosh(\eta-Y) \cosh \eta_t(r) - K_t \frac{x}{r}
\sinh \eta_t(r)\\
\eta_t(r) & = & \eta_f(r/R).
\end{eqnarray*}

\noindent
In this model $(x,y,z,t)$ is the freeze-out position four vector,
$r=\sqrt{x^2+y^2}$ and $\beta_l$ and $M_t$ are the longitudinal
velocity and transverse mass, respectively.  $K_t$ is the transverse
momentum and $\eta_f$ is the transverse flow velocity.  Furthermore, $\Delta
\tau$, $\Delta \eta$ and $R$ describe the respective Gaussian widths
of the proper time, longitudinal velocity and one dimensional Gaussian 
radius.  For more information on the source function, see
\cite{chapman,wiedemann}.


To determine deuteron yields with this framework, a program devised by 
R. Mattiello \cite{mattiello,rqmd108} was implemented.  It utilizes a
method that has been
applied to bombarding energies from 1 GeV per nucleon
in association with the the intranuclear cascade model and QMD
\cite{low_coalesce} up to AGS and SPS energies with the
cascade models RQMD \cite{mattiello,rqmd108} and ARC \cite{kahana}.
In this approach, the number of deuterons is given by a summation over
all proton and neutron pairs at freeze-out
accounting for the Wigner density of the Hulth\'en deuteron
wave function \cite{nagle}.  This method is equivalent to 
the density matrix formalism of Sato and Yazaki \cite{sato} in the
case of a static source.
Therefore, by projecting the neutron/proton position-momentum
distribution on the deuteron wave-function via
this Wigner method, we
may determine the distribution of deuterons in phase-space given any
particular source of nucleons.

\begin{figure}
\centerline{\psfig{figure=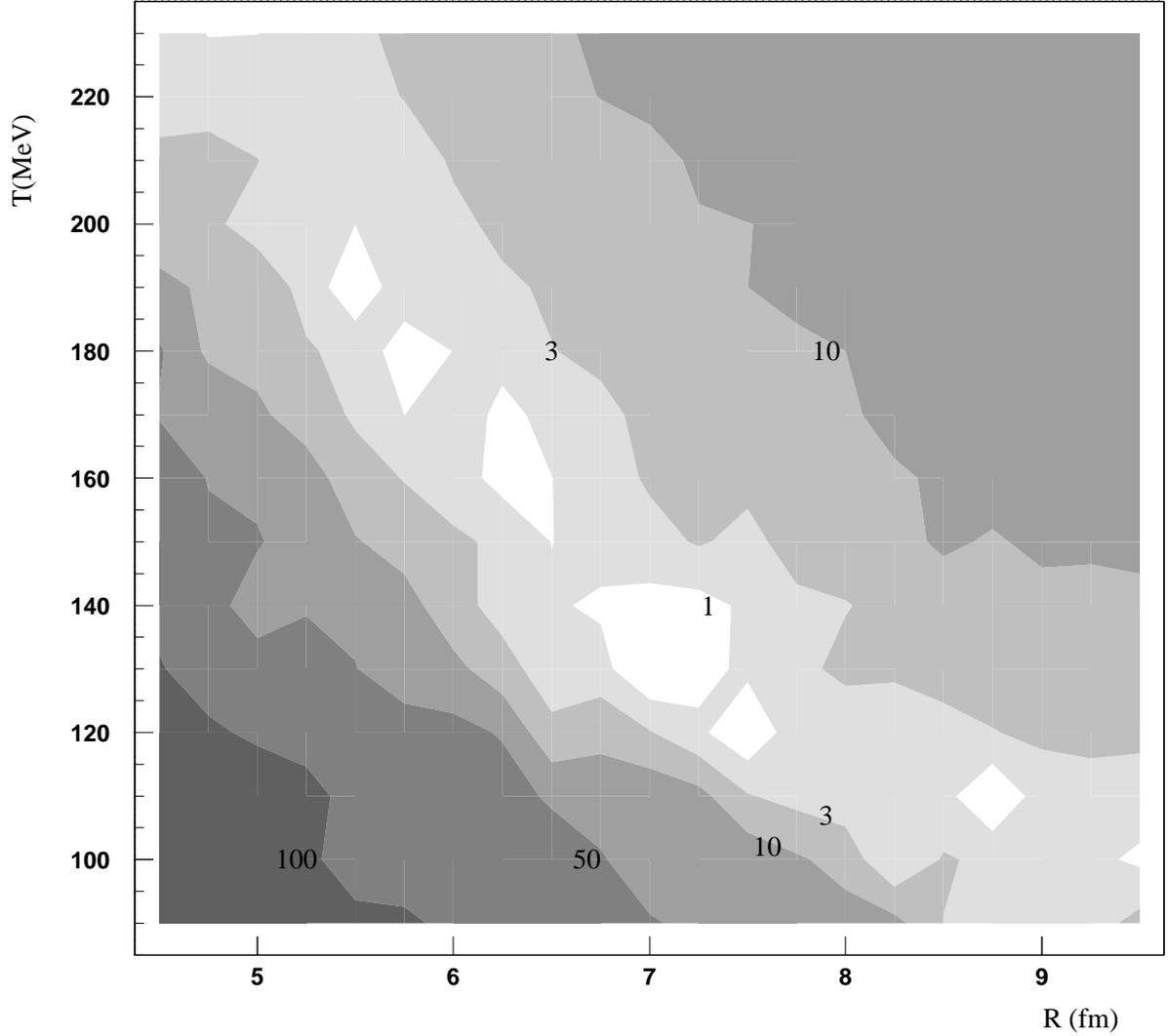,height=17cm}}
\caption{$\chi^2$ map for a fit to an expanding thermal source model
as a function of one-dimensional sideways radius parameter, R,
and temperature, T.  Contours are located at 
$\chi^2 - \chi^2_{min}$ = 1, 3, 10, 50, and 100 and the darker regions 
correspond to larger $\chi^2 - \chi^2_{min}$.  The minimum $\chi^2$ is 
$\chi^2_{min} = 0.96$.  For details see text.}
\label{fig:finally}
\end{figure}

Shown in
Figure~\ref{fig:finally} is the distribution of the reduced $\chi^2$
minus the minimum $\chi^2$ ($\chi^2_{min} = 0.96$) with
respect to the measured deuteron yields calculated as a function of $T$ and
$R$ with a resolution of 10 MeV and 0.25 fm respectively.
For this study we assume that $\tau_0 = 3$fm/c and $\Delta \tau =
1$fm/c in Eqn.~\ref{eqn:source}.
\footnote{For SPS energies, $\Delta \tau \sim 1.5 - 3.0$
fm/c from two particle correlations \cite{wiedemann,na49} while 
\cite{heinz} noted that $\tau_0 \sim 9$fm/c.  Similar
  numbers for AGS energies have not been published.}
The technique we utilize for mapping out this $\chi^2$ space is the
following:  For each value of temperature $T$ and one dimensional
Gaussian radius $R$, the transverse and longitudinal proton spectra
uniquely define the transverse ($\eta_f$) and longitudinal ($\Delta \eta$)
boost velocities.  Every value of $T$ and $R$ in
Figure~\ref{fig:finally} can describe the proton transverse momentum
spectra given a particular $\eta_f$ at mid-rapidity.

We have found that the proton
rapidity distribution in this model poorly reproduces the
data \cite{piazza,866_prot}, with a 
reduced $\chi^2$ on the order of 10 for all choices of $T$, $R$ and
$\Delta \eta$.  Furthermore, the shape of the transverse momentum
distribution of nucleons in this model is assumed to be constant as a
function of rapidity, in conflict with experimental data
\cite{piazza}.  The assumptions inherent in the
model \cite{chapman} regarding a hydrodynamical source are not
entirely valid for nucleons at AGS energies due to these complications 
in the longitudinal direction.  Therefore, in the studies presented
here we only consider the nucleon production at mid rapidity and
implement this model by fitting $\eta_f$ to the proton $p_t$
distribution only at
mid-rapidity and constrain $\Delta \eta$ to approximate the rapidity
distribution.  The deficiencies in the model near beam-rapidity
should not influence the results we show here for mid-rapidity and
indeed the model describes the transverse momentum distribution at
mid-rapidity quite well ($\chi^2 \approx 1$).

For this analysis, the proton spectra from
E866 Au+Au collisions \cite{866_prot} are utilized due to their coverage of
protons at mid-rapidity.  The model is therefore fit with these two
parameters to the proton spectra.  After the proton distribution is
determined, the deuteron yields at mid-rapidity
are calculated with the coalescence algorithm described above.
Results are compared to deuterons measured by E877
at mid-rapidity and $p_t<0.6$ GeV/c.

The resulting $\chi^2$ distribution strongly constrains the
temperature and transverse radius in this model of the nucleon source
though the
two variables are highly correlated.  Studies of proton correlations 
for the Au+Au system at AGS energies \cite{panitkin} concluded there
was an increase of the transverse
freeze-out distribution of nucleons beyond the Au rms radius of
3.1fm.  This is consistent with the current results where all
reasonable values of temperature
result in a value of the transverse radius significantly larger than
the rms radius of a Au nucleus.  Furthermore, measures of the
temperature at freeze-out in Au+Au collisions at the AGS
\cite{thermal,thermal2} of 120-150 MeV are also consistent with the
results of this study.


Note, however, that this method has not constrained
$\Delta \tau$ or $\tau_0$.  Varying these attributes from the values
chosen above by a factor of two can vary the yield of deuterons by 10\%.
An increase in either $\Delta \tau$ or $\tau_0$ would decrease the
yield of deuterons in the model and, as a result, would shift the
$\chi^2$ curve of Figure~\ref{fig:finally} to lower temperatures and
radii.  Therefore, deuterons cannot be used to exclusively determine
source parameters but must be used in conjunction with complimentary
observables to produce a coherent view of the source.


\section*{Conclusions}

We have measured light fragment yields at
both mid and beam rapidities.  Invariant multiplicities
have been presented for deuterons, $^3$H, and $^3$He at mid-rapidity.
Deuterons at mid-rapidity have been interpreted in a coalescence
framework.  The resulting coalescence parameter ($B_d$) is consistent
with an increase in source size from collisions of lighter nuclei at
the same energy.  However, a clear dependence on rapidity of $B_d$
implies the assumption of a constant effective volume as a function of 
rapidity is incorrect.

We have introduced a technique for using deuterons as a sensitive
constraint on
source parameters in a heavy ion collision by incorporating a simple,
hydrodynamically motivated nucleon source function at mid-rapidity
coupled with a coalescence algorithm.  The results show that deuteron
yields are sensitive to a
convolution of volume and collective effects, similar to other
correlations measurements.  These results can be used in conjunction
with other measures of the source to tightly constrain theoretical
models and assumptions.

First measurements of beam rapidity deuterons and $^4$He at AGS
energies have also been made.  The resulting distributions have a
strong $E_t$ dependence, consistent with a production mechanism
dominated by spectator fragmentation.

\section*{Acknowledgements}

We thank the BNL AGS and tandem operations staff and
Dr.~H.~Brown for providing the beam.  This work was supported in part
by the U.S.~DoE, the NSF, the NSERC, Canada and CNPq, Brazil.

We are grateful to Dr.~R. Mattiello, who provided
the coalescence code utilized in this study and to
Drs.~J.~H. Lee and C.~Chasman for
providing us with the results of proton distribution measurements made 
by the E866 collaboration.


\end{document}